\documentclass[twocolumn,showpacs,preprintnumbers,amsmath,amssymb,superscriptaddress]{revtex4}

\usepackage{graphicx}
\begin{document}
\bibliographystyle{prsty}
\title{
Critical test for Altshuler-Aronov theory: Evolution of the density of states singularity in 
double perovskite Sr$_2$FeMoO$_6$ with controlled disorder
}

\author{M. Kobayashi}
\affiliation{Department of Physics and Department of Complexity 
Science and Engineering, University of Tokyo, 
Kashiwa, Chiba 277-8561, Japan}
\author{K. Tanaka}
\affiliation{Department of Applied Physics and 
Stanford Synchrotron Radiation Laboratory, 
Stanford University, Stanford, California 94305, USA}
\author{A. Fujimori}
\affiliation{Department of Physics and Department of Complexity 
Science and Engineering, University of Tokyo, 
Kashiwa, Chiba 277-8561, Japan}
\author{Sugata Ray}
\affiliation{Materials and Structures Laboratory, 
Tokyo Institute of Technology, Midori, 
Yokohama 226-8503, Japan}
\affiliation{Solid State and Structural Chemistry Unit, 
Indian Institute of Science, Bangalore 560012, India}
\author{D. D. Sarma}
\affiliation{Solid State and Structural Chemistry Unit, 
Indian Institute of Science, Bangalore 560012, India}
\affiliation{Centra for Advanced Materials, Indian Association for the Cultivation of Science, Kolkata 700032, India}
\date{\today}

\begin{abstract}
With high-resolution photoemission spectroscopy measurements, the density of states (DOS) near the Fermi level ($E_\mathrm{F}$) of double perovskite Sr$_2$FeMoO$_6$ having different degrees of Fe/Mo antisite disorder has been investigated with varying temperature. The DOS near $E_\mathrm{F}$ showed a systematic depletion with increasing degree of disorder, and recovered with increasing temperature. Altshuler-Aronov (AA) theory of disordered metals well explains the dependences of the experimental results. Scaling analysis of the spectra provides experimental indication for the functional form of the AA DOS singularity.

\end{abstract}

\pacs{71.20.-b, 71.23.-k, 71.27.+a, 79.60.-i}

\maketitle
Disordered electronic systems, which have random potentials deviating from an ideal crystal, have been investigated from both fundamental and application points of view \cite{DES}. 
Ever since the finding of filling-control metal-insulator transitions (MIT) in transition-metal oxides known as strongly correlated system, disorder has attracted even more attention because not only electron-electron interaction but also disorder are supposed to play fundamentally important roles in the MIT. Altshuler and Aronov \cite{Altshuler-Aronov} studied the effect of electron-electron interaction in a disordered metallic medium, and predicted that the density of states (DOS) near the Fermi level ($E_\mathrm{F}$) shows a singularity of $|E-E_\mathrm{F}|^{1/2}$ and the DOS at $E_\mathrm{F}$ increases with increasing temperature in proportion to $\sqrt{T}$. 
The theory has been applied to the low temperature conductivity of disordered metals such as disordered Au and Ag films \cite{Schmitz}, amorphous alloy Ge$_{1-x}$Au$_x$ \cite{GeAu}, and transition metal chalcogenide Ni(S,Se)$_2$ \cite{Honig}.

In a previous work, Sarma {\it et al.} \cite{LNMO} have reported photoemission (PES) measurements on B-site disordered perovskites LaNi$_{1-x}M_x$O$_3$ ($M$=Mn and Fe), which show MIT as a function of $x$, and shown that the disorder affects the DOS near $E_\mathrm{F}$ in such a way that had been theoretically predicted by Altshuler and Aronov \cite{Altshuler-Aronov}. 
In a similar B-site substituted transition-metal oxide SrRu$_{1-x}$Ti$_x$O$_3$, which demonstrates MIT at $x \sim 0.3$ (SrRuO$_3$ is metallic), the depletion of the DOS near $E_\mathrm{F}$ has shown an unusual $|E-E_\mathrm{F}|^{1.2}$ dependence in both metallic and insulating phases \cite{SRTO}. 
In addition, although it is believed that a disorder-induced insulator shows a soft Coulomb gap characterized by a $(E - E_\mathrm{F})^2$ dependence of the DOS near $E_\mathrm{F}$ \cite{Efros, Massey}, the unexpected $|E - E_\mathrm{F}|^{3/2}$ dependence of the DOS near $E_\mathrm{F}$ related to charge density wave has been observed in insulating BaIrO$_3$ \cite{BaIrO3}. 
It is considered that fine structure in the DOS in the vicinity of $E_\mathrm{F}$ is sensitive to both the degrees of disorder and electron correlation, and therefore experimental confirmation of a basic theory for disordered electronic system such as the Altshuler-Aronov (AA) theory is necessary for understanding of the DOS singularity. 
While AA theory makes specific predictions about both $E$ and $T$ dependences, photoelectron spectroscopy has been used only to probe the $E$ dependence with absolutely no reference to the $T$ dependence. Therefore, detailed high-resolution temperature-dependent PES measurements are also highly desired to verify the AA theory.

The present paper reports on high-resolution PES experiments on the B-site ordered double perovskite Sr$_2$FeMoO$_6$ (SFMO), where we have controlled the degree of Fe/Mo antisite disorder (AD) in the sample preparation procedure. Through detailed analysis for the temperature and degree of disorder dependences of the PES spectra near $E_\mathrm{F}$, the results provide experimental confirmation of the AA theory of disordered metals. 
SFMO has been investigated intensively due to the theoretical prediction of half-metallic nature and the observation of large magnetoresistance under low magnetic fields at room temperature \cite{K.-I.Kobayashi}. 
In this system, there are characteristic defects known as Fe/Mo AD at the B-site, which remarkably affects the physical properties of SFMO \cite{SarmaSSC, MRinSFMO, Navarro, B.J.Park, Y.H.Huang}. By controlling the degree of AD, one can investigate disorder effects in the metal without changing the chemical composition and other conditions.

Polycrystalline SFMO samples having different degrees of Fe/Mo AD prepared as follows; 
First, the samples with the highest degree of AD 45\% 
were prepared. Then they were annealed at 1173 K, 1673 K, and 1523 K for a period of 5 hours under 2\% 
H$_2$/Ar to obtain the degrees of AD 40\%, 25\%, and 10\%, respectively (SFMO having AD 50\% 
is the same as ordinary perovskite SrFe$_{0.5}$Mo$_{0.5}$O$_3$). Details of the sample preparation are given in Ref.~\cite{SarmaSSC} and \cite{MRinSFMO}. Using x-ray diffraction, the degree of disorder was quantified from the intensity of a supercell-reflection peak. Scanning electron microscopy in conjunction with energy dispersive x-ray analysis revealed no change in composition during the annealing. 
Transport measurements were performed on the AD 10\% and 40\% samples by a standard four prove technique using a Physical Property Measurement System (Quantum Design Co. Ltd). 
PES spectra were recorded using a spectrometer equipped with a monochromatized He resonance lamp ($h{\nu}=21.2$ eV), where photoelectrons were collected with a Gammadata Scienta SES-100 hemispherical analyzer in the angle integrated mode. The total resolution of the spectrometer was ${\sim}10$ meV, and the base pressure was $1.0{\times}10^{-8}$ Pa. Clean surfaces were obtained by repeated scraping {\it in situ} with a diamond file. The position of $E_\mathrm{F}$ was determined by measuring PES spectra of evaporated gold which was electrically in contact with the sample.

\begin{figure}[!t]
\begin{center}
\includegraphics[width=8.85cm]{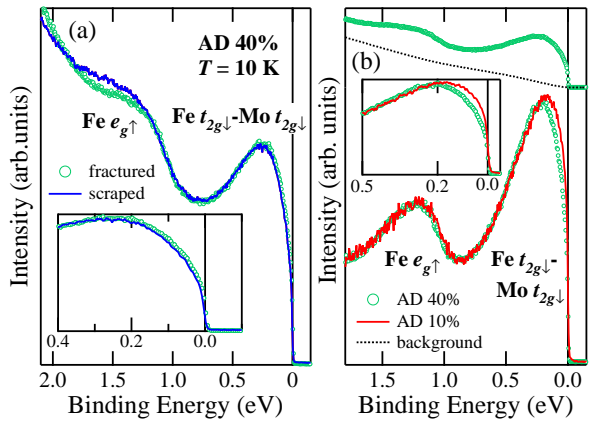}
\caption{
Valence-band photoemission spectra of Sr$_2$FeMoO$_6$ taken at 10 K with He-I radiation. 
(a) Comparison between the spectra taken from the fractured and scraped surfaces. 
The spectra have been normalized to the area from 0 eV to 0.8 eV (Fe $t_{2g\downarrow}$-Mo $t_{2g\downarrow}$ states). 
(b) Comparison between different degrees of antisite disorder (AD) 10\% and 40\%. 
The spectra have been normalized to the area from 0.8 eV to 2 eV (Fe $e_{g\uparrow}$ states). 
Top: Background subtraction. The insets show enlarged plots near $E_\mathrm{F}$.}
\label{NarrowScan}
\end{center}
\end{figure}


The line shapes of the PES spectra obtained were almost the same as those in a previous report on single crystalline SFMO taken with synchrotron radiation \cite{T.Saitoh}. In order to examine the influence of surface treatment, we measured valence-band spectra taken from fractured and scraped surfaces. In the binding-energy ($E_B$) range from 2 eV to 10 eV, there were appreciable differences such as the sharpness of structures in the O~2$p$ and the Fe~$t_{2g\uparrow}$ bands (not shown), the background intensity, and to some extent the Fe~$e_{g\uparrow}$ band. On the other hand, within $\sim1$ eV of $E_\mathrm{F}$, i.e., in the Fe~$t_{2g\downarrow}$ $+$ Mo~$t_{2g\downarrow}$ conduction bands, the line shapes of the fractured and scraped samples were similar to each other as shown in Fig.~\ref{NarrowScan}(a). 
Since in previous PES studies it has been reported that LDA+$U$ calculation well explains the valence-band spectra taken from the fractured surface \cite{T.Saitoh, Bulk-PES}, we consider that the different surface treatments have not affected the spectra near $E_\mathrm{F}$ which reflect the bulk properties. In contrast, the spectra are intensively influenced by AD and temperature as we shall see below.

\begin{figure}[!t]
\begin{center}
\includegraphics[width=8.95cm]{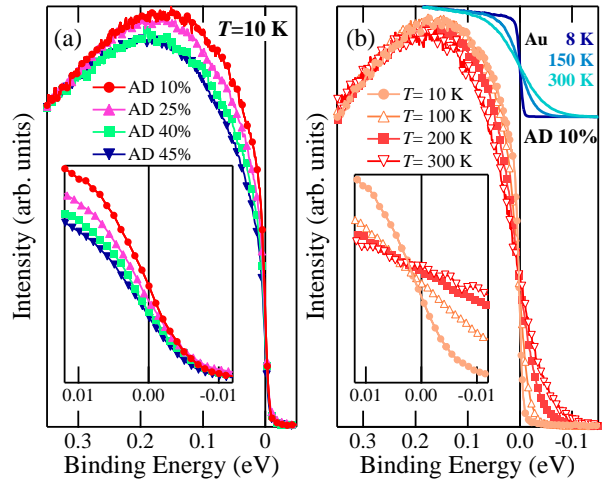}
\caption{
Photoemission spectra of Sr$_2$FeMoO$_6$ near the Fermi level. 
The spectra have been normalized to the area from $E_B=0.3$ eV to 0.6 eV. 
(a) Degree of Fe/Mo AD dependence at 10 K. 
(b) Temperature dependence of the AD 10\% sample. 
As a reference, Au spectra are also shown. The insets show an enlarged plot in the vicinity of $E_\mathrm{F}$.
}
\label{ADandTemp}
\end{center}
\end{figure}

Figure~\ref{NarrowScan}(b) shows valence-band spectra for different degrees of disorder, i.e., AD 40\% 
and 10\%. 
Although the peak due to the localized Fe $e_{g\uparrow}$ states was nearly identical between AD 10\% 
and AD 40\%, 
there was a clear difference in the Fe $t_{2g \downarrow}$-Mo $t_{2g\downarrow}$ conduction band between the two spectra. This result suggests that the disorder influences the DOS near $E_\mathrm{F}$. 
Figure~\ref{ADandTemp} shows the temperature and degree of disorder dependences of the spectra near $E_\mathrm{F}$ normalized to the area in the region from $E_B = 0.3$ eV to 0.6 eV, in which the spectra were identical and independent of the degree of AD as shown in Fig.~\ref{NarrowScan}(b). 
For a fixed temperature, the intensity of the spectra near $E_\mathrm{F}$ was depleted with degree of disorder. This behavior is consistent with the previous report on a disordered metal system LaNi$_{1-x}$Mn$_x$O$_3$ \cite{LNMO}. 
Indeed, temperature dependent resistivity measurements on the AD 40\% 
sample showed a minimum around 40 K, i.e., the resistivity increased with decreasing temperature below 40 K, while on the AD 10\% 
one there was no minimum till the lowest measured temperature (20 K). The observations are consistent with previous reports of transport measurements on SFMO having various degrees of disorder \cite{Y.H.Huang, Y.H.Huang2} and on SFMO with high Fe/Mo ordering \cite{MRinSFMO, SingleCrystalSFMO}. 
For a fixed degree of disorder, the intensity at $E_\mathrm{F}$ increased with temperature as shown in Fig.~\ref{ADandTemp}(b). This behavior indicates that SFMO differs from normal metals such as Au in which PES spectra at various temperatures have temperature-independent intensity at $E_\mathrm{F}$ and intersect at $E_\mathrm{F}$ irrespective of temperature as shown in Fig~\ref{ADandTemp}(b), representing the simple Fermi-Dirac distribution function.

Now, we analyze the temperature and degree of disorder dependences of the spectra based on the theory for disordered metal suggested by Altshuler and Aronov \cite{Altshuler-Aronov}. The theory predicted that electron-electron interaction accompanied by impurity scattering leads to an anomaly in the DOS around $E_\mathrm{F}$ and the resulting singular part of the DOS ${\delta}{\mathcal{D}}$ is given by
\begin{equation}
\begin{split}
\label{AA}
\frac{{\delta}{\mathcal{D}}({\epsilon})}{{\mathcal{D}}_0} &= 
 \frac{4{\pi}e^2}{\mathcal{H}^2}\frac{1}{2\sqrt{2}{\pi}^2} 
  \frac{\sqrt{k_\mathrm{B}T}}{({\hbar}D)^{3/2}}
   {\varphi} \left( {|\epsilon|}/{k_\mathrm{B}T} \right), \\
{\varphi}&\left( X \right) =\left\{
	\begin{split}
	\sqrt{X} \qquad\quad\  &(X \gg 1) \\
	1.07 + O(X^2) \quad &(X \ll 1)
	\end{split}
\right.,
\end{split}
\end{equation}
where ${\mathcal{D}}_0$ is the original DOS at $E_\mathrm{F}$, $\epsilon$ is the energy measured from $E_\mathrm{F}$, $\mathcal{H}$ is the inverse Debye radius, and $D$ is the diffusion coefficient due to impurity scattering.


\begin{figure}[!t]
\begin{center}
\includegraphics[width=8.8cm]{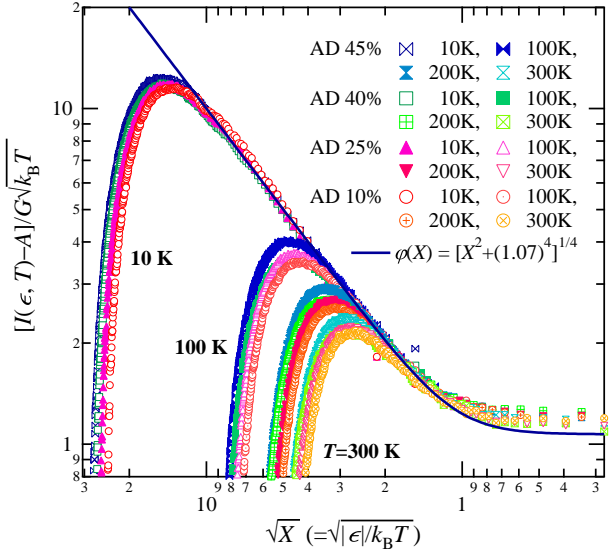}
\caption{
Scaling analysis for the spectral depletion. 
Scaled spectra as a function of $\sqrt{X}$ ($X=|\epsilon|/k_\mathrm{B}T$) are plotted on a bilogarithmic scale. 
An analytical form of $\varphi=[X^2 + (1.07)^4]^{1/4}$ is also plotted. 
}
\label{Scaling}
\end{center}
\end{figure}

In order to see whether experimental spectra satisfy Eq.~(\ref{AA}), we have made the following scaling analysis. According to Eq.~(\ref{AA}), the PES spectra $I(\epsilon, T)$ near $E_\mathrm{F}$ should be proportional to $\mathcal{D}_0' + \delta\mathcal{D}(\epsilon)$, where $\mathcal{D}_0'$ is a constant and $\mathcal{D}_0' < \mathcal{D}_0$. Therefore, $I(\epsilon, T)$ can be parameterized as 
\begin{equation}
\label{Intensity}
I(\epsilon, T) = A + G\, \sqrt{k_\mathrm{B}T}\, \varphi \left(\frac{|\epsilon|}{k_\mathrm{B}T} \right),
\end{equation}
where $A$ and $G$ are only dependent on the degree of disorder. 
$(I-A)/G\sqrt{k_\mathrm{B}T}$ plotted against $X=|\epsilon|/k_\mathrm{B}T$ should fall onto the same curve if Eq.~(\ref{AA}) is valid, and can be used to evaluate the functional form of $\varphi$ if $A$ and $G$ are chosen to satisfy the condition that $\varphi(0) \to 1.07$ as $\epsilon/k_\mathrm{B}T \to 0$. In Fig.~\ref{Scaling}, the results are plotted on a logarithmic scale, where the PES spectra have been divided by the Fermi-Dirac function convoluted with the experimental resolution estimated from the Au spectra. 
Notice that all the low energy part of the spectra fell onto the same curve, which we attribute to the scaling function $\varphi(X)$. 
Deviation from the scaling function $\varphi(X)$ occurs at high energies where the original DOS starts to deviate from the constant one. 
We find that $\varphi(X)$ approaches $\sqrt{X}$ for $X>1$, corresponding to AA theory. 
The results ensure the validness of analysis based on AA theory for the depletion of PES spectra near $E_\mathrm{F}$.


\begin{figure}[!t]
\begin{center}
\includegraphics[width=8.8cm]{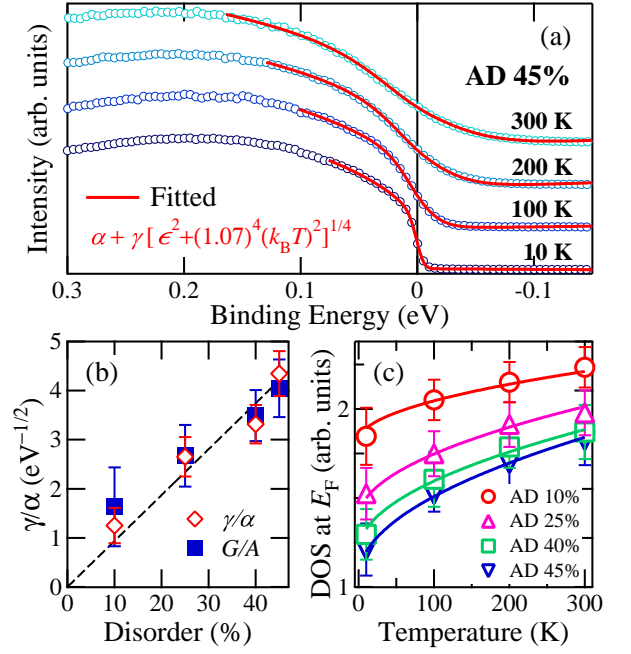}
\caption{
Results of fitting of the photoemission spectra of Sr$_2$FeMoO$_6$ to the analytical curves of Altshuler-Aronov theory. 
(a) Fitted spectra for the AD 45\% sample at various temperatures. 
(b) Degree of disorder dependence of the values of $\gamma/\alpha$. $G/A$ is also plotted. The dotted line is a guide to the eye. 
(c) Temperature dependence of the DOS at $E_\mathrm{F}$. Solid curves demonstrate fitted curves using Eq.~(\ref{root}) with parameters deduced from the spectral line-shape fitting of panel (a). 
}
\label{FittingResults}
\end{center}
\end{figure}

In order to analyze the spectra using Eq.~(\ref{AA}), we propose an analytical form of $\varphi( X )=[X^2 + (1.07)^4]^{1/4}$ interpolating both the limits of large and small $X$ of Eq.~(\ref{AA}). 
This form is shown to accurately reproduce the experimentally scaling function $\varphi(X)$ deduced above as shown in Fig.~\ref{Scaling}. Therefore, we employ a model function 
\begin{equation}
\{ \alpha + \gamma \, [\epsilon^2 + (1.07)^4 (k_\mathrm{B}T)^2]^{1/4} \} \, f(\epsilon),
\label{fitting}
\end{equation}
where $f(\epsilon)$ is the Fermi-Dirac function, $\alpha$ and $\gamma$ are fitting parameters. $\gamma/\alpha$ depends only on degree of disorder and are independent of a way of intensity normalization. 
Figure~\ref{FittingResults}(a) shows fitted results for the spectra of the AD 45\% 
sample at various temperatures. The fitting function given by Eq.~(\ref{fitting}) well reproduced the spectra near $E_\mathrm{F}$, where the fitted ranges were chosen to the valid range of the scaling function $\varphi(X)$ as shown in Fig.~\ref{Scaling} \cite{SPofDev}. 
Figure~\ref{FittingResults}(b) shows the values of the coefficients which represent the strength of the DOS singularity as a function of disorder. The $\gamma/\alpha$ value is independent of temperature and approximately linearly increases with degree of disorder as shown in Fig.~\ref{FittingResults}(b), indicating that the DOS singularity near $E_\mathrm{F}$ is enhanced with increasing degree of disorder as predicted by AA theory. The constant $G$ will be relative to the value of $\gamma$. Actually, the $G/A$ value shows the same dependence as $\gamma/\alpha$ [Fig.~\ref{FittingResults}(b)]. 
Equation~(\ref{fitting}) as a function of $X$ without $f$, i.e., $\alpha + \gamma[X^2 + (1.07)^4]^{1/4}$, well reproduced the line shape of the depletion as shown in Fig.~\ref{Scaling}, where the parameters were chosen $\alpha=0$ and $\gamma=1$ to correspond with the analytical $\varphi(X)$ \cite{ExponentN}. 
The result demonstrates validness of our assumption for the functional form of $\varphi$ given by Eq.~(\ref{fitting}).

Equation~(\ref{AA}) indicates that the singular contribution to the DOS $\delta\mathcal{D} / {\mathcal{D}_0}$ at $E_\mathrm{F}$ is proportional to $\sqrt{T}$. 
In order to study the temperature dependence of the DOS at $E_\mathrm{F}$, comparison was made between the experimental and theoretical $\delta\mathcal{D} /{\mathcal{D}_0}$ at $E_\mathrm{F}$. 
The PES intensity (or DOS) at $E_\mathrm{F}$ is plotted as a function of temperature in Fig.~\ref{FittingResults}(c). For a fixed temperature, the DOS at $E_\mathrm{F}$ increased with decreasing degree of disorder. For a fixed degree of disorder, the DOS at $E_\mathrm{F}$ increased with increasing temperature. According to Eq.~(\ref{AA}), temperature dependence of the DOS at $E_\mathrm{F}$ is expressed as 
\begin{equation}
\label{root}
\alpha+ 1.07\, \gamma\sqrt{k_\mathrm{B}T},
\end{equation}
corresponding Eq.~(\ref{fitting}) at $\epsilon = 0$. 
Using the values of $\gamma/\alpha$ obtained by fitting to the PES spectra, Eq.~(\ref{root}) well reproduces the temperature-dependence of the DOS at $E_\mathrm{F}$ as shown in Fig.~\ref{FittingResults}(c). The result is consistent with theoretically predicted temperature dependence of $\delta \mathcal{D}/\mathcal{D}_0$ at $E_\mathrm{F}$. 
It follows from the arguments described above that AA theory applies to not only the disorder dependent depletion near $E_\mathrm{F}$ but also the temperature dependent DOS at $E_\mathrm{F}$.

As mentioned above, AA theory treats the scattering processes on general ground and does not depend on the functional form of the potential of the scattering center. This theory can be applied to the case that the mean free path of itinerant electrons is larger than or comparable with its wave length.

In conclusion, we have performed high-resolution photoemission experiments on polycrystalline Sr$_2$FeMoO$_6$ samples having different degrees of Fe/Mo antisite disorder. The photoemission spectra near the Fermi level depended on degree of the Fe/Mo antisite disorder as well as on temperature. The Altshuler-Aronov theory on disordered metal well explained both the dependences. Scaling analysis for the spectral depletion clarifies the functional form of the density of states singularity near the Fermi level. 
We believe that the findings will provide an indicator for degrees of disorder and electron-electron correlation, and promote spectral analysis for the index of the density of states depletion near the Fermi level. 
The present results point to a need for taking into account both electron-electron interaction and disorder effects for an understanding of the electronic structure of metallic correlated electron system.

The authors thank H. Yagi and M. Hashimoto for help in experiments. This work was supported by a Grant-in-Aid for Scientific Research in Priority Area ``Invention of Anomalous Quantum Materials'' (16076208) from MEXT, Japan. D.D.S. thanks DST and BRNS for funding this research. S.R. thanks JSPS postdoctoral fellowship for foreign researchers. M.K. acknowledges support from the Japan Society for the Promotion of Science for Young Scientists.


\begin{thebibliography}{10}
\bibitem{SPofDev}
The starting points are about 76 meV, 109 meV, 129 meV, and 145 meV for 10 K,
  100 K, 200 K, and 300 K, respectively.
\bibitem{ExponentN}
The analytical form of $\varphi(X)$ can be expressed as more general formula:
  $\varphi_n( X ) = [X^{2n} + (1.07)^{4n}]^{1/4n}$, which also satisfies both
  the limits of $X$. Using this function, the scaling function has been fitted
  well as $n=1.09 \pm 0.14$. This result emphasizes the validness of the
  analytical $\varphi(X)$ because the exponent $n$ is nearly 1.
\end{thebibliography}
\end{document}